# Anomalous Scaling Exponents of Capacitance –Voltage Characteristics

Vikas Nandal and Pradeep R. Nair

Department of Electrical Engineering, Indian Institute of Technology Bombay, Mumbai, Maharashtra, India-400076

Email: nknandal@iitb.ac.in, prnair@ee.iitb.ac.in

*Abstract-* **Capacitance-Voltage (CV) measurements along with the Mott-Schottky (MS) analysis are widely used for characterization of material and device parameters. Using a simple analytical model, supported by detailed numerical simulations, here we predict that the capacitance of thin film devices scale as $V^{-2}$ (V is the applied potential), instead of the often used $V^{-0.5}$ dependence of MS analysis – with significant implications towards extraction of parameters like doping density, built-in voltage, etc. Surprisingly, we find that the predicted trends are already hidden in multiple instances of existing literature. As such, our results constitute a fundamental contribution to semiconductor device physics and are directly applicable and immediately relevant to a broad range of optoelectronic devices like organic solar cells, perovskite based solar cells and LEDs, thin film a-Si devices, etc.**

Organic and hybrid materials based optoelectronic devices have attracted immense research focus due to their low processing cost,[1,2] high flexibility,[2–6] and excellent optical properties.[7–9] Material and device characterization is one of the crucial steps towards identification of functional parameters that dictate the performance of electronic devices. Although widely used, the MS analysis,[10] in this regard is strictly valid only for $P^+N$ or $PN^+$ device architectures. Novel optoelectronic devices such as organic solar cells,[11,12] perovskite based solar cells[13–16] and LEDs,[17–19] thin film a-Si devices[20,21] often employ carrier selective PIN based architectures.[1,22–24] However, the capacitance-voltage characteristics of such carrier selective PIN devices is not very well elucidated in the literature. Still, the MS analysis is widely used for such devices which could result in potentially erroneous extraction of critical parameters like the built-in voltage, effective doping density, and thus could negatively affect the co-optimization of fabrication and device architecture towards better performance. Therefore, it is imperative to build a theoretical platform to understand the CV characteristics of thin film PIN based optoelectronics devices.

In this manuscript, we develop an analytical model for the CV characteristics of carrier selective PIN devices. In particular, we find that the capacitance is governed by the injected charge carriers from the respective contacts and scales as $V^{-2}$ (in contrast to traditional $V^{-0.5}$ relation). Interestingly, these analytical predictions are well supported by detailed numerical simulations. In addition, curiously, we find multiple instances of direct experimental evidence from literature which broadly supports the analytical predictions. Below, we first develop an analytical formalism for the CV of carrier selective devices and then validate it through numerical simulations and experimental data.

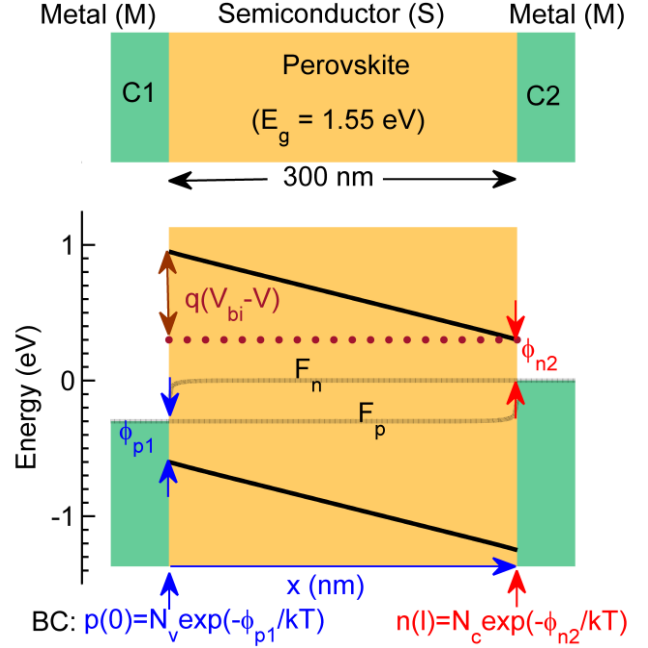

**Figure 1**. Device schematic and energy band diagram of the MSM model system, with perovskite as the test material, at an applied voltage $V = 0.3\,V$ at metal contact C1. Quasi-fermi level ($F_n, F_p$) and electric field in active (Perovskite) layer is constant with position in the low bias regime (for $V < V_{bi}$) of carrier transport. Electrons and holes are injected from contact C2 and C1, respectively, inside perovskite with the applied voltage V. Electron density $n_2 = n(l)$ (at Perovskite/C2 interface) and hole density $p_1 = p(0)$ (at C1/Perovskite interface) are limited by electron ($\phi_{n2}$) and hole ($\phi_{p1}$) barrier, respectively.

**Analytical Model:** The capacitance of a semiconductor device is a measure of response of charge carriers to a superimposed small sinusoidal signal (of a certain frequency) over a particular DC voltage.[10] As such, it is then expected that the CV of PIN devices to be influenced by (a) the depletion region in P and N regions, and (b) the charge stored in the intrinsic as well as quasi-neutral layers. The influence of (a) on CV characteristics is rather well anticipated by the MS analysis while that of (b) is not well appreciated. To this end, we first consider a metal-semiconductor-metal (MSM) system such that the charge storage component due to the intrinsic layer is well elucidated. The analysis will then be extended to carrier selective PIN structures as well. The schematic of MSM systems is shown in Fig. 1. Here a semiconductor is sandwiched between two metal contacts C1 and C2. The workfunction of these contacts are so chosen that C1 injects holes while C2 injects electrons to the semiconductor, respectively. For MSM, the total capacitance is the parallel combination of geometric capacitance $C_g$ and charge storage capacitance $C_d$ (i.e., rate of change of stored charge with applied voltage). Hence, to develop CV model, we calculate $C_d$ by quantifying the total stored charge (due to electrons and holes) at a particular voltage. For electrons, the transport in drift-diffusion formalism is given as

$$J_n = qn\mu_n E + qD_n \frac{dn}{dx}, \quad (1)$$

where $J_n$ is electron current density, $n$ is electron density, $\mu_n$ is electron mobility, $E$ is electric field, $D_n = \mu_n kT/q$ is diffusion constant for electrons, and $x$ is the location/position inside semiconductor. Under the assumption of constant electric field in the I layer, the electron density profile as a function of position $n(x)$ can be calculated by rearranging Eq. 1 and integrating both sides with appropriate boundary condition ($n(l) = n_2$). Hence, we obtain electron density profile $n(x)$ as follows

$$n(x) = \frac{J_n}{q\mu_n E} - \left(\frac{J_n}{q\mu_n E} - n_2\right) \exp\left(-\frac{qE(x-l)}{kT}\right). \quad (2)$$

Here $k$ is Boltzmann's constant, $T$ is the device temperature, $n_2$ is the electron density injected from contact C2 at $x = l$, $l$ is the thickness of semiconducting layer, and $E = (V_{bi} - V)/l$ ($V_{bi}$ is the built-in voltage and $V$ is the applied bias). The total stored electron charge ($n_T$), at a particular voltage $V$, can be obtained by integrating $n(x)$ across the thickness of semiconducting layer. Accordingly, for low bias regime (i.e. $V < V_{bi}$), we get a simplified expression of $n_T$ in terms of V as

$$n_T = \frac{n_2 lkT}{q(V_{bi} - V)}, \quad (3)$$

Replacing $n_2 = N_c \exp(-\phi_{n2}/kT)$, where, $N_c$ is conduction band effective density of states and $\phi_{n2}$ is injection barrier for electrons from metal contact C2 to semiconductor, the stored charge capacitance due to electrons ($C_{n,d} = q\frac{dn_T}{dV}$) is given as

$$C_{n,d} = \frac{N_c \exp\left(-\frac{\phi_{n2}}{kT}\right) lkT}{(V_{bi} - V)^2}, \quad (4)$$

Similar methodology can be used to obtain stored charge capacitance due to the response of holes ($C_{p,d}$) and is expressed as

$$C_{p,d} = \frac{N_v \exp\left(-\frac{\phi_{p1}}{kT}\right) lkT}{(V_{bi} - V)^2}. \quad (5)$$

Here, $N_v$ is valence band effective density of states and $\phi_{p1}$ is injection barrier for holes from metal contact C1 to semiconductor. The total capacitance $C_d$ is due to the response of electrons and holes, and therefore, is the parallel combination of $C_{n,d}$ and $C_{p,d}$ (i.e. $C_d = C_{n,d} + C_{p,d}$). In contrast to traditional linearity between $1/C^2$ and $V$, indeed, our model (shown in Eq. 4, 5) predicts linear relation between $1/C^{0.5}$ and $V$ and hence can be used to estimate $V_{bi}$. For asymmetrical injection barriers (i.e. $\phi_{n2} \neq \phi_{p1}$), the same method allows estimation of the properties of the contact that dominates carrier injection – which can be further probed through temperature dependent studies of CV characteristics.

**Model Validation:** We performed detailed numerical simulations to validate the analytical model. For this, we consider intrinsic perovskite (with $E_g$=1.55 eV and thickness $l = 300\ nm$) as the test material with $\phi_{n2} = \phi_{p1} = 0.3\ eV$ (see Fig. 1). Small signal analysis using self-consistent solution of drift-diffusion and Poisson's equation is used to obtain the CV characteristics.

Fig. 2a shows the comparison of electron density variation with applied voltage as obtained from simulations and analytical expressions (given in Eq. 2). The results indicate that, in low bias regime ($V < V_{bi}$), the electron density reduces exponentially with the distance from respective contacts. The integrated/total charge density variation with applied voltage is shown in Fig. 2b. It is evident from Fig. 2a and 2b, in low bias regime (shown by green region in Fig. 2b), that the proposed model agrees very well with the numerical simulations.

Fig. 2c shows the simulated CV characteristics of the modeled MSM device. We find that the capacitance saturates to geometrical capacitance $C_g = \epsilon/l$ in negative biasing regime. Here, $\epsilon$ and $l$ is the electrical permittivity and thickness of semiconducting layer, respectively. For $0 < V < V_{bi}$, the capacitance increases with voltage as the stored charge in perovskite layer increases (as evident from Fig. 2a and 2b). For $V > V_{bi}$, the MSM device becomes dominated by semiconductor resistance and hence the capacitance decreases after reaching a peak. We can extract the stored charge capacitance component as $C_d = C - C_g$, where $C$ is the total capacitance. Now, as suggested by CV model in Eq. 4 and 5, one can predict $V_{bi}$ and the injected charge carriers density ($n_2$ and $p_1$) from the intercept and slope of $1/C_d^{0.5}$ vs. $V$ plot, respectively. We find that the extracted carrier density ($4.09 \times 10^{14} cm^{-3}$) closely matches with the input values of injected charge density ($n_2 = p_1 = 9 \times 10^{14} cm^{-3}$). In addition, Fig. 2d shows that the $V_{bi}$ predictions from our model are in close agreement with the ideal values for a broad range of $V_{bi}$. In contrast, MS analysis does not provide accurate results for $V_{bi}$ and hence is not an appropriate characterization scheme for such devices.

We further validated the CV model by extracting barrier for carrier injection from CV characteristics and comparing it with the input value. For this, we estimated CV characteristics (numerically) at different temperatures ($T = 300 - 380\ K$). The CV characteristics due to injected charge carriers after $C_g$ correction (i.e. $C_d = C - C_g$) are plotted in Fig. 2e. With the increase in temperature, the flux of injected charge carriers from respective contacts increases and thus the total charge stored inside perovskite layer increases. As a result, the capacitance increases with the rise in temperature at a particular voltage (Fig. 2e). Equations 4 and 5 indicate that $C_d \propto NkT \exp\left(-\frac{\phi}{kT}\right)$, which is of the form of well-known Arrhenius equation ($y \propto \exp(-x/kT)$) except that it has temperature dependent prefactor $NkT$. Here we consider that the effective density of states $N \propto T^{3/2}$ (as generally expected in semiconductors[10]) and hence the prefactor $NkT \propto T^{5/2}$. Correction of such temperature dependent prefactor is essential to obtain accurate results for $\phi$ from CV characteristics. Accordingly, Fig. 2f shows Arrhenius plots at different voltages with $\log(C_d/T^{5/2})$ as y axis and $1/kT$ as x axis. We extract $\phi$ from the magnitude of the slope of linear fits made on $\log(C_d/T^{5/2})$ vs. $1/kT$ plots (obtained at different voltages). We find that the extracted value of $\phi$ is in close agreement with the input value of 0.3 eV. Our simulations indicate that the linearity in $1/C_d^{0.5}$ vs. $V$ characteristics is influenced by the presence of semiconductor doping/bulk traps. Detailed numerical simulations in this regard and CV characteristic of PIN devices are provided in supplementary materials.

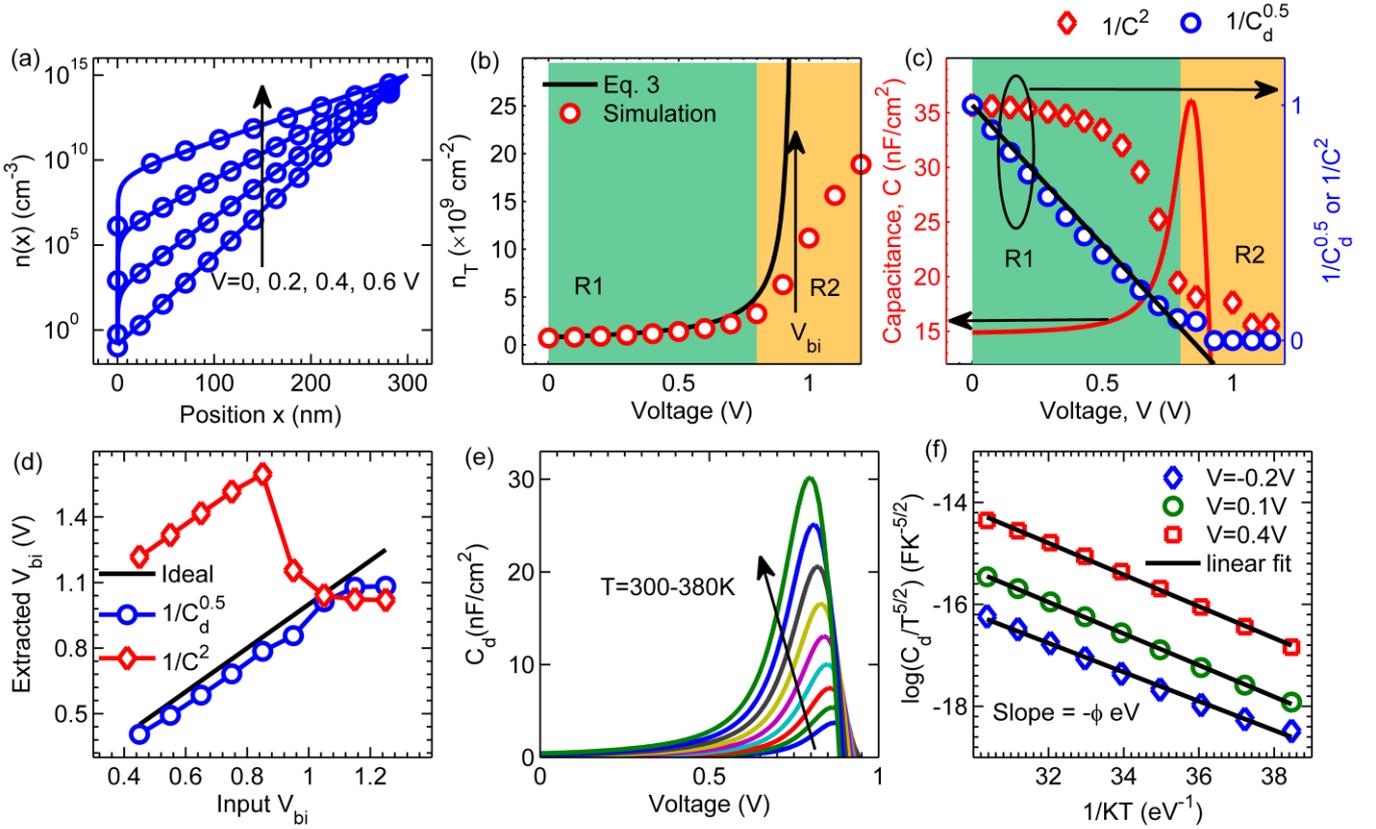

**Figure 2**. Analytical model validation of CV characteristics by detailed numerical simulations of the MSM device. Comparison between analytical (line) and simulated (circle symbols) (a) electron density $n(x)$ at low voltages ($V < V_{bi}$, arrow is in the direction of voltage increase) (b) integrated electron $n_T$ for the given range of applied voltages. (c) Simulated CV characteristic is plotted on the left axis, whereas, $1/C_d^{0.5}$ or $1/C^2$ vs. $V$ (with symbols) is plotted on the right axis. Single linear fit (shown by black line) made on $1/C_d^{0.5}$ vs. $V$ plot suggest that our model is valid and is used to extract relevant parameters such as $V_{bi}$. (d) Comparison between ideal and extracted $V_{bi}$ (using $1/C^2$ vs. $V$ and $1/C_d^{0.5}$ vs. $V$ plot) suggests that our model ($C \propto V^{-2}$) provides good estimates of $V_{bi}$ as compared to traditional model ($C \propto V^{-0.5}$) for broad range of $V_{bi}$. (e) $C_d - V$ characteristics at different temperature $T$ (T=300 K to 380 K, in steps of 10 K and aligned with arrow direction). (f) Arrhenius plot at voltages $V$ (= $-0.2, 0.1, 0.4$ V) where y axis represents $\log(C_d/T^{5/2})$ and x axis corresponds to $1/kT$. The magnitude of slope of linear fit made on Arrhenius plot provides injection barrier $\phi$ at the contacts. The extracted barrier is independent of voltage and matches well with the input (0.3 eV). Analytical CV Model is well supported by numerical simulations in low bias regime (represented by green region R1 shown in panel b and c) and hence can be used to predict relevant parameters ($V_{bi}$, interface charge density).

**Experimental validation:** So far, we showed that the proposed CV model is a convenient tool to extract the relevant parameters ($V_{bi}, n_2/p_1, \phi$) and the results are very well supported by numerical simulations. Given the simplicity of the analytical predictions and its close correlations with numerical simulations, we expect the trends from the analytical model to be reproduced in broad experimental results as well. To further investigate the claims of the analytical model, we collected several recent experimental data[25–30] from literature (with different PIN based technologies and reported from multiple laboratories). The CV characteristics of those results are then plotted as $1/C^2$ vs. $V$ (Mott-Schottky plot) and $1/C_d^{0.5}$ vs. $V$ in Fig. 3. Here $C_d = C - C_g$, where $C_g$ is the geometric capacitance and can be considered as the lowest value of capacitance in reverse bias regime.

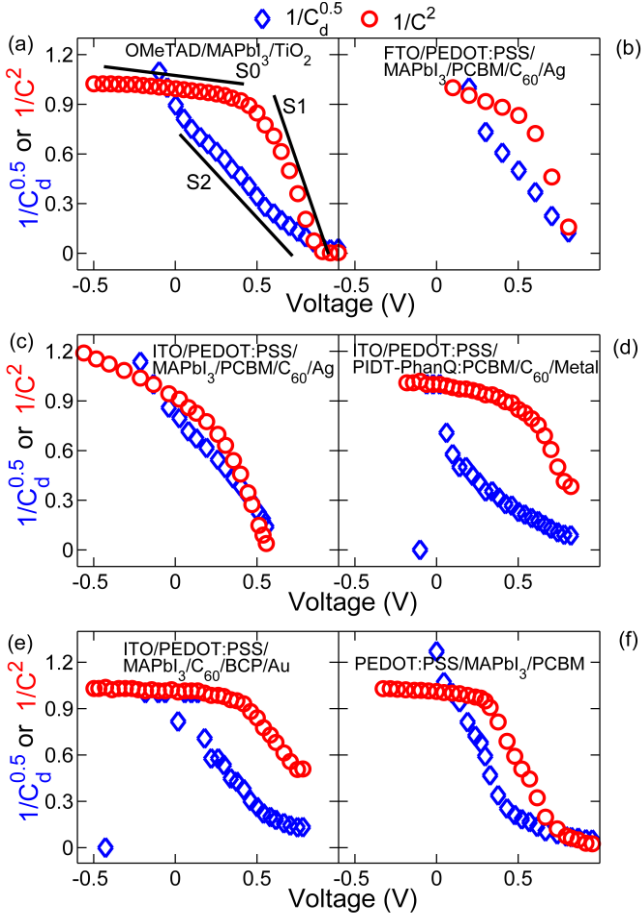

**Figure 3.** Experimental validation of the proposed model (i.e., $C_d \propto V^{-2}$, see equations 4 and 5). CV characteristics of six different experimental devices (a- ref 25; b-ref 26; c-ref 27; d- ref 28; e- ref 29; f- ref 30) are represented in terms of $1/C^2 - V$ (Mott-Schottky, shown as red circles) and $1/C_d^{0.5} - V$ (shown as blue diamonds) plots. Note that each curve is normalized to the respective value at V=0. Mott-Schottky plots exhibit anomalous behavior with two different linear regimes, however, $1/C_d^{0.5} - V$ plots exhibit single linear regime over a broad range of voltages – thus validating the analytical model.

The results in Fig. 3 show several interesting trends. First, almost all of them shows linear trends in $1/C_d^{0.5} - V$ characteristics in the low bias regime – a conclusive supporting evidence for the analytical model presented in this work. Further, the results in Fig. 3 indicate that Mott-Schottky plots are, at the best, piece wise linear over rather small voltage ranges. Such a behavior indicates that the subsequent parameter extraction could be suspicious and in general the usage of MS analysis should be avoided. In contrast, the linearity in low bias regime of $1/C_d^{0.5} - V$ characteristics extends over a broad range of voltage and hence allows unambiguous extraction of device parameters. These independent experimental results conclusively settle the debate of the voltage scaling exponents of CV characteristics. It is indeed $V^{-0.5}$ (as we proposed) and not $V^{-2}$ (as understood so far in the community). The experimental support is remarkable given the fact the analytical model was developed for simple MSM structures while the experimental results were obtained using carrier selective PIN structures (see Supplementary materials for numerical simulations on CV of PIN structures). Further, we notice that for certain devices (see, Fig. 3c, e, &f), both $1/C_d^{0.5} - V$ and $1/C^2$ vs. $V$ exhibit almost similar slopes. In the next section, through detailed numerical simulations indicate, we show that such a puzzling trend could arise due to the presence of unintentional doping or bulk traps.

**Effects of doping/trap density**: Until now, we considered the CV characteristics of PIN device in the absence traps or dopants in the active material (see Figure S2, supplementary information). However, in practical devices, there could be scenarios where the active material contains defect states and/or dopant atoms. Therefore, it is imperative to incorporate doping or trap density into the model system and understand their effects on the linearity of $1/C_d^{0.5}$ vs. $V$ plot. For this, we consider two cases: (a) with acceptor type trap density $N_t^{AL}$ present at single energy level $E_t$ (with respect to valence band energy $E_v$), and e (b) with acceptor doping density $N_a^{AL}$ in active material. Detailed numerical simulations, with the incorporation of traps or doping density in PIN based device, are performed to obtain CV characteristics which are then plotted as $1/C_d^{0.5}$ vs $V$ and $1/C^2$ vs. $V$ in figure 4. The material parameters used for numerical simulations are provided in Table S2 (see supplementary information).

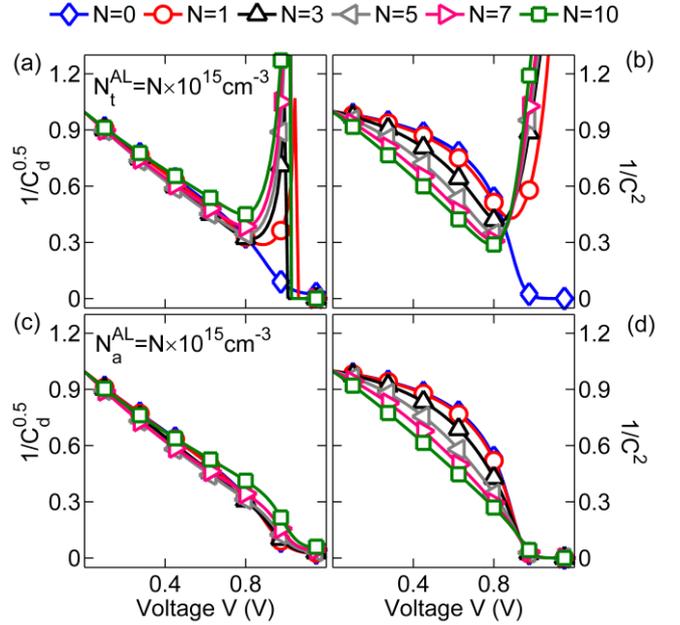

**Figure 4.** Influence of bulk traps/doping density on the linearity of $1/C_d^{0.5}$ and $1/C^2$ with voltage $V$. CV characteristics of PIN devices are plotted as $1/C_d^{0.5}$ vs. $V$ (in left panels) and $1/C^2$ vs. $V$ (in right panels) at different values of bulk trap ($N_t^{AL}$ – panel a, b)/doping ($N_a^{AL}$ – panel c, d) density. The results indicate that, in low bias regime, $1/C^2$ vs. $V$ plot changes from non-linear to linear with the increase in bulk trap/doping density, whereas, $1/C_d^{0.5}$ vs. $V$ plot remains linear for the given range of trap/doping density. Hence, our analytical model ($C \propto V^{-2}$) is valid for devices having given range of bulk traps/doping density, however, the traditional model

($C \propto V^{-0.5}$) is only applicable for devices with high bulk trap/doping density.

The results in figure 4 shows the effects of bulk trap/doping density on the curvature of $1/C_d^{0.5}$ vs $V$ and $1/C^2$ vs. $V$ (MS analysis) plots. Interestingly, we find that the behavior of $1/C^2$ with applied voltage $V$ changes from non-linear to linear with increase in bulk trap density. This is mainly caused by the electrostatic effects (energy band profile changes from linear to non-linear) due to trapped charge carriers or ionized dopant atoms in the active material. Surprisingly, in contrast to MS analysis, the linear nature of $1/C_d^{0.5}$ vs. $V$ plot is not significantly affected by the presence of bulk traps/dopant atoms. Indeed, our model ($C \propto V^{-2}$) is valid for wide variety of PIN based devices having different bulk traps/dopant atoms, whereas, the traditional model is valid for large values of trap/doping density. Further, similar to experimental trends in Figure 3, we notice that the slopes of CV characteristics (i.e., $1/C_d^{0.5}$ vs $V$ and $1/C^2$ vs. $V$) are nearly equal for certain bulk trap/doping densities. Such observations regarding slopes in different plots can provide the evidence of bulk trap/doping density in the active material. Hence, careful examination of degree of linearity in respective plots is imperative to obtain accurate estimation of concerned parameters from the analysis of CV characteristics.

**Conclusions**: To summarize, we provided a theoretical platform with physical insights that could predict the scaling exponents of capacitance voltage characteristics of carrier selective PIN based optoelectronic devices. Our analytical results indicate that CV characteristics show linearity between $1/C^{0.5}$ and $V$ in contrast to the traditional Mott-Schottky analysis. Our CV model is well supported by detailed numerical simulation and by diverse experimental data. Based on these results, we recommend that every experimental report on C-V characteristics of thin film semiconductor devices should display both $1/C_d^{0.5} - V$ and $1/C^2$ vs. $V$ for appropriate parameter extraction. As such this study is relevant and influential for a broad class of optoelectronic devices. As such this study is relevant and influential for a broad class of optoelectronic devices with immediate relevance to experimentalists.

ACKNOWLEDGEMENT

This article is based upon work supported under the US-India Partnership to Advance Clean Energy-Research (PACE-R) for the Solar Energy Research Institute for India and the United States (SERIIUS), funded jointly by the U.S. Department of Energy (Office of Science, Office of Basic Energy Sciences, and Energy Efficiency and Renewable Energy, Solar Energy Technology Program, under Subcontract DE-AC36-08GO28308 to the National Renewable Energy Laboratory, Golden, Colorado) and the Government of India, through the Department of Science and Technology under Subcontract IUSSTF/JCERDC-SERIIUS/2012 dated 22nd November 2012. The authors also acknowledges Center of Excellence in Nanoelectronics (CEN) and National Center for Photovoltaic Research and Education (NCPRE), IIT Bombay for computational facilities.

METHODS

Through self-consistent solution of Poisson's and continuity equation[10] for electrons and holes, detailed numerical simulations were performed[31] to analyze low frequency capacitance voltage characteristics of the modeled system. The details of material parameters used for numerical simulations of MSM and PIN device are provided in table S1 and S2, respectively (see supplementary material).

SUPPORTING INFORMATION

The supporting information is available free of charge on ACS publications website.

The supporting information includes: Material parameters for numerical simulation of MSM and PIN device; Effects of doping/trap density on CV characteristics of MSM device; Validation of CV model for PIN device;

**Notes**

The authors declare no competing financial interest